\documentclass[11pt,headsepline]{scrartcl}

\usepackage{amssymb}
\usepackage{amsfonts}
\usepackage{amsmath}
\usepackage{graphicx}

\DeclareMathSymbol{\C}{\mathalpha}{AMSb}{"43}
\DeclareMathSymbol{\R}{\mathalpha}{AMSb}{"52}
\begin{document}
\newcommand{\mytitle}{A New Case for Direct Action}
\title{\mytitle}

\author{Michael Ibison \\
             Institute for Advanced Studies at Austin,\\
             11855 Research Boulevard, Austin, TX 78759, USA\\
%             \email{ibison@earthtech.org}
             email: ibison@earthtech.org
}

\pagestyle{myheadings} \markright{Michael Ibison \hfill \mytitle}

\date{}
\maketitle

\begin{abstract}
An obstacle to the development of direct action version of electromagnetism was that in the end it failed to fulfill its initial promise of avoiding the problem of infinite Coulomb self-energy in the Maxwell theory of the classical point charge. This paper suggests a small but significant modification of the traditional direct action theory which overcomes that obstacle. Self-action is retained but the associated energy is rendered finite and equal to zero in the special case of null motion.
\end{abstract}

%PACS: 03.50.De; 03.50.Kk; 03.50.-z; 41.90.+e; 11.10.Lm; 12.90.+b

%Keywords: self-energy; self-action; renormalization; direct action

%%%%%%%%%%%%%%%%%%%%%%%%%%%%%%%%%%%%%%%%%%%%%%%%%%%%%%%%%%%%%%%%%%%%%%%%%%%%%%%%%%%%%%%%%%%%%%%%%%%%%%%%%%%%%
\section{Background}
%%%%%%%%%%%%%%%%%%%%%%%%%%%%%%%%%%%%%%%%%%%%%%%%%%%%%%%%%%%%%%%%%%%%%%%%%%%%%%%%%%%%%%%%%%%%%%%%%%%%%%%%%%%%%
Arguably there are three essential characteristics peculiar to EM direct action that distinguish it from the Maxwell field-theoretic approach:
\begin{enumerate}
  \item[i)] The (initial) promise of zero self-energy.
  \item[ii)] EM time symmetry, and hence the need to find an explanation for the observed asymmetry of radiation outside of EM.
  \item[iii)]  No vacuum degrees of freedom, hence no EM second-quantization and no EM ZPF.
\end{enumerate}
The focus below is exclusively on i). Wheeler and Feynman \cite{WF1,WF2} attempted to resolve ii) though it was later found that the cosmological boundary condition - identified as necessary in order to explain the predominance of retarded radiation - could not be met by any reasonable cosmology \cite{Davies3,Davies4}. In any case their argument has since been criticized on the grounds that it hides the assumption of a thermodynamic arrow of time upon which it bases a derivation of the EM arrow of time \cite{Price5,Zeh6}. An alternative suggestion \cite{Ibison7} remains unproven. Davies \cite{Davies8,Davies9,Davies10} has shown that the results of relativistic QM are not affected by iii) provided all radiation is absorbed.\footnote{That which is generally taken to be retarded radiation is then really an interaction between sources.} Boyer \cite{Boyer11} amongst others has shown that some of behaviours attributed to the EM ZPF of QED such as the Lamb shift and the Casimir effect can be treated within stochastic electrodynamics (SED) - a variant of classical EM augmented by the presence of a (classical) Lorentz-Invariant background field.\footnote{The Casimir analysis in Itzykson and Zuber \cite{IZ12} for example is essentially classical and equal to that of SED.}

%%%%%%%%%%%%%%%%%%%%%%%%%%%%%%%%%%%%%%%%%%%%%%%%%%%%%%%%%%%%%%%%%%%%%%%%%%%%%%%%%%%%%%%%%%%%%%%%%%%%%%%%%%%%%
\section{Brief Derivation of Direct Action}
%%%%%%%%%%%%%%%%%%%%%%%%%%%%%%%%%%%%%%%%%%%%%%%%%%%%%%%%%%%%%%%%%%%%%%%%%%%%%%%%%%%%%%%%%%%%%%%%%%%%%%%%%%%%%
The direct action paradigm is that the EM potentials are not independent degrees of freedom apart from the currents that source them. If the field theoretic form of the classical action is taken to be\footnote{$A$ and $j$ are Lorentz vectors with suppressed indexes. The symbol $\circ$ denotes a Lorentz scalar product so that $A \circ j = A^aj_a = \phi \rho - \bf{A . j}$, for example, where $\bf{A.j}$ is the Euclidean scalar product. $dx^2 = dx \circ dx$. $q$ is a particle index. Everywhere $c = 1$.}
\begin{equation} \label{MaxwellAction}
I = \int {d^4 x} \left\{ {\frac{1}
{2}A \circ \left( {\partial ^2 A} \right) + \kappa\left( {\partial  \circ A} \right)^2  - A \circ j} \right\} - \sum\limits_q {m_q \int {\sqrt {dx_q^2 } } }
\end{equation}
then variation of $A$ would normally give $\partial ^2 A = j$ in the Lorenz ($\kappa = 0$) gauge with the implication that\footnote{The symbol $*$ denotes convolution so that $G * j = \int {d^4 x'} G\left( {x - x'} \right)j\left( {x'} \right)$, $\delta$ is the Dirac delta function.}
$\label{convolution}
A = G * j + A_{cf} ;\quad \partial ^2 G = \delta ^4 ,\quad \partial ^2 A_{cf}  = 0
$
for some (scalar) Green's function $G$ appropriate to the boundary conditions. By contrast direct action maintains that $A_{cf}  = 0$ for some $G$ that can be expressed as independent of $j$: no fields exist that cannot be explained by currents. Implementing this in (\ref{MaxwellAction}) gives
\begin{equation} \label{GreensAction1}
I =  - \frac{1}
{2}\int {d^4 x} \,j \circ G * j - m_q \int {\sqrt {dx_q^2 } }\,.
\end{equation}
Explicating the currents as\footnote{$x$ and $x_q \left( \lambda  \right)$ are both 4-vectors. $\lambda$ parameterizes the world line. It is not necessarily a Lorentz scalar. $\delta ^4 \left( {x - x_q \left( \lambda  \right)} \right)$ is shorthand for $\prod\limits_{a = 0}^3 {} \delta \left( {x^a  - x_q^a \left( \lambda  \right)} \right)$ where $a$ indexes the components of a Lorentz vector.}
$$
j\left( x \right) = \sum\limits_q {e_q \int d x_q \left( \lambda  \right)\delta ^4 \left( {x - x_q \left( \lambda  \right)} \right)}
$$
the action becomes\footnote{Notice there is no gauge freedom in (\ref{GreensAction2}).}
\begin{equation} \label{GreensAction2}
I =  - \frac{1}
{2}\sum\limits_{p,q} {e_p e_q \int d x_p \left( \kappa  \right) \circ \int d x_q \left( \lambda  \right)\,} G\left( {x_p \left( \kappa  \right) - x_q \left( \lambda  \right)} \right) - \sum\limits_q {m_q \int {\sqrt {dx_q^2 } } }\,.
\end{equation}
The traditional choice in direct action is the symmetric propagator of advanced and retarded influences $G\left( x \right) = \delta \left( {x^2 } \right)/4\pi$. (Any anti-symmetric component, even if permitted, makes no contribution to the action due to the permutation symmetry of the particles.) With this and explicit parameterization the action becomes
\begin{equation} \label{DirectActionwithMass}
I =  - \frac{1}
{{8\pi }}\sum\limits_{p,q} {e_p e_q \int {d\kappa } \int {d\lambda } \,} \dot x_p \left( \kappa  \right) \circ \dot x_q \left( \lambda  \right)\delta \left( {\left( {x_p \left( \kappa  \right) - x_q \left( \lambda  \right)} \right)^2 } \right) - \sum\limits_q {m_q \int {\sqrt {dx_q^2 } } }\,.
\end{equation}
%
%%%%%%%%%%%%%%%%%%%%%%%%%%%%%%%%%%%%%%%%%%%%%%%%%%%%%%%%%%%%%%%%%%%%%%%%%%%%%%%%%%%%%%%%%%%%%%%%%%%%%%%%%%%%%
\section{Traditional Electromagnetic Self-Energy}
%%%%%%%%%%%%%%%%%%%%%%%%%%%%%%%%%%%%%%%%%%%%%%%%%%%%%%%%%%%%%%%%%%%%%%%%%%%%%%%%%%%%%%%%%%%%%%%%%%%%%%%%%%%%%
The total action (\ref{DirectActionwithMass}) for $N$ particles can be written as
\begin{equation} \label{NparticleAction}
I = \sum\limits_{\scriptstyle p,q \hfill \atop
  \scriptstyle p \ne q \hfill} ^{} {} I_{p,q}  + NI_{self}  - \sum\limits_q^{}m_q  {\int {\sqrt {dx_q^2 } } }
\end{equation}
where $I_{self}$ is the electromagnetic self-action
$$
I_{self}  =  - \frac{{e^2 }}
{{8\pi }}\mathop {\lim }\limits_{\varepsilon  \to 0} \sum\limits_{\sigma  =  \pm 1} {w\left( \sigma  \right)\int {d\kappa } \int {d\lambda } \,\dot x\left( \kappa  \right) \circ \dot x\left( \lambda  \right)\delta \left( {\left( {x\left( \kappa  \right) - x\left( \lambda  \right)} \right)^2  - \sigma \varepsilon ^2 } \right)}
$$
where the Lorentz scalar $\varepsilon$ is a small regularization parameter  - sufficiently small to capture only the local light-cone self-intersections of the world line. $\sigma  =  \pm 1$ determines the time-like or space-like nature of the regularization and $w\left( \sigma  \right)$ is a dimensionless weight of order unity. The limit $\varepsilon  \to 0$ is to be taken after extremization. Performing one of the integrations to extract the contribution from the delta function near $\kappa  = \lambda$ \footnote{$\Theta \left( x \right)$ is the Heaviside step function. Its value exactly at $x=0$ is immaterial here since exactly light-speed motion is excluded.}:
\begin{equation} \label{SelfAction}
\begin{aligned}
  I_{self}  =  &  - \frac{{e^2 }}
{{8\pi }}\mathop {\lim }\limits_{\varepsilon  \to 0} \sum\limits_{\sigma  =  \pm 1} {w\left( \sigma  \right)\int d \xi \int {d\lambda \,} \dot x\left( {\lambda  + \xi } \right) \circ \dot x\left( \lambda  \right)\delta \left( {\left( {x\left( {\lambda  + \xi } \right) - x\left( \lambda  \right)} \right)^2  - \sigma \varepsilon ^2 } \right)}  \\
   =  &  - \frac{{e^2 }}
{{8\pi }}\mathop {\lim }\limits_{\varepsilon  \to 0} \sum\limits_{\sigma  =  \pm 1} {w\left( \sigma  \right)\int {d\lambda \,} \left( {\dot x^2 \left( \lambda  \right) + O\left( \xi  \right)} \right)\int d \xi \delta \left( {\xi ^2 \dot x^2 \left( \lambda  \right) + O\left( {\xi ^3 } \right) - \sigma \varepsilon ^2 } \right)}  \\
   =  &  - \frac{{e^2 }}
{{16\pi }}\mathop {\lim }\limits_{\varepsilon  \to 0} \left( {\sum\limits_{\sigma  =  \pm 1} {w\left( \sigma  \right)\int {d\lambda \,} \operatorname{sgn} \left( {\dot x^2 \left( \lambda  \right)} \right)\left. {\frac{1}
{{\left| \xi  \right|}}} \right|_{_{\xi ^2 \dot x^2 \left( \lambda  \right) = \sigma \varepsilon ^2 } }  + O\left( \xi^0  \right)} } \right)\\
   =  &  - \mathop {\lim }\limits_{\varepsilon  \to 0} \frac{{e^2 }}
{{8\pi \left| \varepsilon  \right|}}\int {d\lambda \,} \left( {w\left( 1 \right)\Theta \left( {\dot x^2 \left( \lambda  \right)} \right)\sqrt {\dot x^2 \left( \lambda  \right)}  - w\left( { - 1} \right)\Theta \left( { - \dot x^2 \left( \lambda  \right)} \right)\sqrt { - \dot x^2 \left( \lambda  \right)} } \right) \\
&+  O\left( \varepsilon^0  \right)\,.
\end{aligned}
\end{equation}
Note that the above is invalid if the world line is exactly null. Contribution to the action at order $O\left( {\varepsilon ^0 } \right)$ is ignored since the primary interest here is to render the self-energy finite in the limit. Specializing in this paper to time-monotonic world lines the integration parameter can be set to the laboratory time\footnote{Time reversals, which necessitate that at least some part of the trajectory is spacelike, were implicitly excluded at the outset by the particular form of the mechanical action. They are not excluded by the electromagnetic part of the action alone however.} $\lambda  = x_0 \left( \lambda  \right) = t$
\begin{equation} \label{SelfDensity}
\begin{aligned}
  I_{self}  =  & \int {dt\,} L_{self} \left( {{\mathbf{v}}\left( t \right)} \right) \\
  L_{_{self} }  =  &  - \mathop {\lim }\limits_{\varepsilon  \to 0} \frac{{e^2 }}
{{8\pi \left| \varepsilon  \right|}}\left( {w\left( 1 \right)\Theta \left( {1 - {\mathbf{v}}^2 } \right)\sqrt {1 - {\mathbf{v}}^2 }  - w\left( { - 1} \right)\Theta \left( {{\mathbf{v}}^2  - 1} \right)\sqrt {{\mathbf{v}}^2  - 1} } \right) \\
\end{aligned}
\end{equation}
for which the canonical self-energy is\footnote{Use has been made of the condition that the world line is not null, whereupon the derivatives of the Heaviside functions vanish.}
\begin{equation} \label{SelfEnergy}
H_{self}  =  {\bf{v}}{\bf{.}}\frac{{\partial L_{_{self} } }}
{{\partial {\bf{v}}}} - L_{_{self} }  = \mathop {\lim }\limits_{\varepsilon  \to 0} \frac{{e^2 }}
{{8\pi \left| \varepsilon  \right|}}\left[ {w\left( 1 \right)\frac{{\Theta \left( {1 - {\bf{v}}^2 } \right)}}
{{\sqrt {1 - {\bf{v}}^2 } }} + w\left( { - 1} \right)\frac{{\Theta \left( {{\bf{v}}^2  - 1} \right)}}
{{\sqrt {{\bf{v}}^2  - 1} }}} \right]
\end{equation}
provided ${\bf{v}}^2  \ne 1$. The usual singular result for the electromagnetic self-energy is now evident as $\varepsilon  \to 0$.

In traditional classical electrodynamics the classical charged particle is confined to the sub-luminal domain with a finite rest mass achieved through cancellation of the infinite electromagnetic self-energy by the mechanical energy $m_{bare} /\sqrt {1 - {\bf{v}}^2 }$, the latter coming from the third term in (\ref{NparticleAction}).\footnote{A second contribution to mass renormalization comes from the ZPF in QED and stochastic electrodynamics.} The corresponding regularizing factors are $w\left( 1 \right) = 1$ and $w\left( { - 1} \right) = 0$. The method outlined here is a Lorentz-Invariant generalization of the familiar method employing a small charged shell whose radius goes to zero.\footnote{Notice in this derivation the absence of a problem associated with a factor of 4/3 destroying Lorentz covariance  - see \cite{Hnizdo13} for a nice review of the history of that problem.} The energy (\ref{SelfEnergy}) as a function of speed is given in figure 1.
\begin{figure}[htp]
\centering
\includegraphics[width=0.8\textwidth]{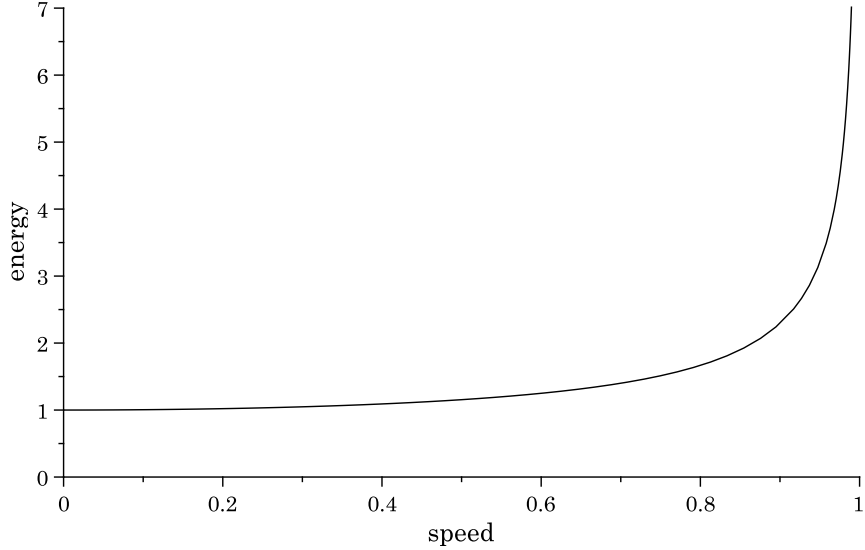}
\caption{Regularized Maxwell self-energy as function of speed. All magnitudes $\rightarrow \infty$ as $\varepsilon\rightarrow 0$.}\label{fig:1}
\end{figure}

%%%%%%%%%%%%%%%%%%%%%%%%%%%%%%%%%%%%%%%%%%%%%%%%%%%%%%%%%%%%%%%%%%%%%%%%%%%%%%%%%%%%%%%%%%%%%%%%%%%%%%%%%%%%%
\section{Finite self-energy}
%%%%%%%%%%%%%%%%%%%%%%%%%%%%%%%%%%%%%%%%%%%%%%%%%%%%%%%%%%%%%%%%%%%%%%%%%%%%%%%%%%%%%%%%%%%%%%%%%%%%%%%%%%%%%
An early attraction of direct action was that it permitted the easy elimination of self energy by excluding self-action from the action Schwarzschild \cite{STF14}, Tetrode \cite{STF15}, and Fokker \cite{STF16}. In (\ref{DirectActionwithMass}) $\smash{\sum_{p,q}  \to  \sum_{p,q;\;p \ne q}}$ and therefore $\smash{H_{_{self} }  = L_{_{self} }  = 0}$. Later it was realized that self-action generates physically observable consequences and so cannot be simply removed - at least from QED. Feynman \cite{Feynman17} for example noted that pair creation can be regarded as a form of promoted self-action. One of the alleged advantages of direct action (over field theory) was then lost.

An alternative evident from (\ref{SelfEnergy}) is to choose the regularization $w\left( \sigma  \right) = \sigma$, so the total action is
\begin{equation} \label{weightedAction}
\begin{aligned}
I =  &- \frac{1}
{{8\pi }}\sum\limits_{\sigma  =  \pm 1} {\sum\limits_{p,q} {e_p e_q w\left( \sigma  \right)\int {d\kappa } \int {d\lambda } \,} } \dot x_p \left( \kappa  \right) \circ \dot x_q \left( \lambda  \right)\delta \left( {\left( {x_p \left( \kappa  \right) - x_q \left( \lambda  \right)} \right)^2  - \sigma \varepsilon ^2 } \right)
\\ &- \sum\limits_q {m_q \int {\sqrt {dx_q^2 } } }\,,
\end{aligned}
\end{equation}
the self action is
\begin{equation} \label{finiteSelfAction}
L_{_{self} }  =  - \mathop {\lim }\limits_{\varepsilon  \to 0} \frac{{e^2 }}
{{8\pi \left| \varepsilon  \right|}}\left( {\Theta \left( {1 - {\bf{v}}^2 } \right)\sqrt {1 - {\bf{v}}^2 }  + \Theta \left( {{\bf{v}}^2  - 1} \right)\sqrt {{\bf{v}}^2  - 1} } \right)
\end{equation}
and the self-energy
\begin{equation} \label{finiteSelfEnergy}
H_{self}  = {\bf{v}}{\bf{.}}\frac{{\partial L_{_{self} } }}
{{\partial {\bf{v}}}} - L_{_{self} }  = \mathop {\lim }\limits_{\varepsilon  \to 0} \frac{{e^2 }}
{{8\pi \left| \varepsilon  \right|}}\left[ {\frac{{\Theta \left( {1 - {\bf{v}}^2 } \right)}}
{{\sqrt {1 - {\bf{v}}^2 } }} - \frac{{\Theta \left( {{\bf{v}}^2  - 1} \right)}}
{{\sqrt {{\bf{v}}^2  - 1} }}} \right]
\end{equation}
is plotted in figure 2. This implies the energy is zero at light speed and is infinite in magnitude at all other speeds (positive infinite at sub-luminal speeds, negative infinite at superluminal speeds). The steps leading to (\ref{SelfEnergy}) are not valid precisely at light speed however. An interpretation consistent with (\ref{finiteSelfAction}) is that the charge is comprised of two zero-dimensional parts infinitesimally separated in space and moving respectively infinitesimally faster and slower than the speed of light ($\left| {\bf{v}} \right| = 1_ +  , 1_ -$, say).
\begin{figure}[htp]
\centering
\includegraphics[width=0.8\textwidth]{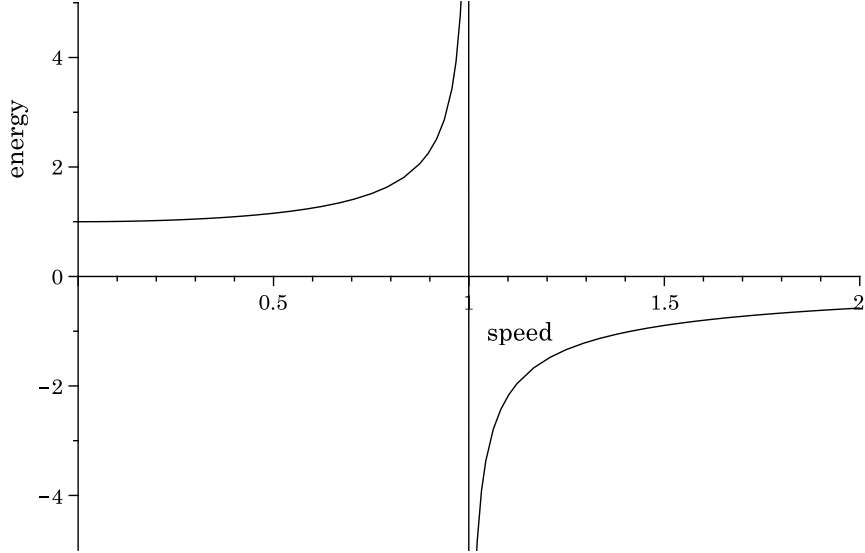}
\caption{Regularized self-energy of charge due to modified direct action. All non-zero magnitudes $\rightarrow \infty$ as $\varepsilon\rightarrow 0$.}\label{fig:2}
\end{figure}

Another alternative reported previously \cite{Ibison175} and having the same end result in this case\footnote{The two approaches are not generally equivalent, specifically in the entirely superluminal domain away from the nearly null motion considered here.} is to modify the action in (\ref{DirectActionwithMass}) so that the scalar product $\dot x_p \left( \kappa  \right) \circ \dot x_q \left( \lambda  \right)$ is replaced by the positive definite $\left| {\dot x_p \left( \kappa  \right) \circ \dot x_q \left( \lambda  \right)} \right|$ - which change cannot affect the predictions of the action in the sub-luminal domain. Then the regularized version of (\ref{DirectActionwithMass}) can be written
\begin{equation}
\begin{aligned}
  I =  &  - \frac{1}
{{8\pi }}\sum\limits_{\sigma  =  \pm 1} {\sum\limits_{p,q} {e_p e_q w\left( \sigma  \right)\int {d\kappa } \int {d\lambda } \,} } \left| {\dot x_p \left( \kappa  \right) \circ \dot x_q \left( \lambda  \right)} \right|\delta \left( {\left( {x_p \left( \kappa  \right) - x_q \left( \lambda  \right)} \right)^2  - \sigma \varepsilon ^2 } \right) \\
      &  - \sum\limits_q {m_q \int {\sqrt {dx_q^2 } } }  \\
   =  &  - \frac{1}
{{8\pi }}\sum\limits_{\sigma  =  \pm 1} {\sum\limits_{p,q} {e_p e_q w\left( \sigma  \right)\int{d\kappa } \int {d\lambda } \,} } \delta \left( {\frac{{\left( {x_p \left( \kappa  \right) - x_q \left( \lambda  \right)} \right)^2  - \sigma \varepsilon ^2 }}
{{\dot x_p \left( \kappa  \right) \circ \dot x_q \left( \lambda  \right)}}} \right) \\
      &  - \sum\limits_q {m_q \int {\sqrt {dx_q^2 } } }\,.
\end{aligned}
\end{equation}
Following the steps in (\ref{SelfAction}) the modified self action is then
\begin{equation}
\begin{aligned}
  I_{self}  =  &  - \frac{{e^2 }}
{{8\pi }}\mathop {\lim }\limits_{\varepsilon  \to 0} \sum\limits_{\sigma  =  \pm 1} {w\left( \sigma  \right)\int {d\xi } \int {d\lambda } \,\delta \left( {\frac{{\left( {x\left( {\lambda  + \xi } \right) - x\left( \lambda  \right)} \right)^2  - \sigma \varepsilon ^2 }}
{{\dot x\left( {\lambda  + \xi } \right) \circ \dot x\left( \lambda  \right)}}} \right)}  \\
   =  &  - \mathop {\lim }\limits_{\varepsilon  \to 0} \frac{{e^2 }}
{{8\pi \left| \varepsilon  \right|}}\int {d\lambda \,} \left( {w\left( 1 \right)\Theta \left( {\dot x^2 \left( \lambda  \right)} \right)\sqrt {\dot x^2 \left( \lambda  \right)}  + w\left( { - 1} \right)\Theta \left( { - \dot x^2 \left( \lambda  \right)} \right)\sqrt { - \dot x^2 \left( \lambda  \right)} } \right)
\end{aligned}
\end{equation}
so the effect is just to change the sign of $w\left( { - 1} \right)$ from how it appears in (\ref{SelfDensity}). The modified energy is then the same as (\ref{finiteSelfEnergy}) provided now $w\left( \sigma  \right) = 1$ (figure 2).

Since the mechanical mass is no longer required for Coulomb mass renormalization and in any case is infinite at light speed it can now be dropped altogether from the (modified) classical theory, in which case any finite observed mass must here be attributed entirely to an external interaction in the spirit of Higgs. The simplest possibility is that the external mass-giving interaction is entirely electromagnetic necessitating therefore some sort of EM ZPF i.e. at zero Kelvin.

%%%%%%%%%%%%%%%%%%%%%%%%%%%%%%%%%%%%%%%%%%%%%%%%%%%%%%%%%%%%%%%%%%%%%%%%%%%%%%%%%%%%%%%%%%%%%%%%%%%%%%%%%%%%%
\section{Effective local action}
%%%%%%%%%%%%%%%%%%%%%%%%%%%%%%%%%%%%%%%%%%%%%%%%%%%%%%%%%%%%%%%%%%%%%%%%%%%%%%%%%%%%%%%%%%%%%%%%%%%%%%%%%%%%%
As written the self-action (\ref{finiteSelfAction}) is unsatisfactory; the limit cannot be taken in advance of extremization, and whatever finite non-zero contributions may remain as $\varepsilon \rightarrow 0$ have already been discarded in computing (\ref{SelfAction}). In lieu of a fuller treatment an effective Lagrangian is sought without these shortcomings and which captures the quality of (\ref{finiteSelfAction}) that enforces light speed motion, from which finite external forces are capable of causing only infinitesimal departures. A candidate is
\begin{equation} \label{effectiveAction}
\begin{aligned}
I_{self}  =  &- \frac{1}
{2}\int {d\lambda \,} \mu \left( \lambda  \right)\dot x^2 \left( \lambda  \right) - \int {d^4 x\,} A^{\left(ext\right)}  \circ j \\
=  &- \int {d\lambda \,} \dot x\left( \lambda  \right) \circ \left( {\frac{1}
{2}\mu \left( \lambda  \right)\dot x\left( \lambda  \right) + eA^{\left(ext\right)} \left( {x\left( \lambda  \right)} \right)} \right)
\end{aligned}
\end{equation}
where $\smash{A^{\left(ext\right)}}$ is the potential due to non-self sources. $\mu$ is an undetermined multiplier variation of which generates the light speed constraint $\dot x^2 \left( \lambda  \right) = 0$. Variation of $x\left( \lambda  \right)$ gives
\begin{equation} \label{NewtonLorentz}
\frac{d}{{d\lambda }}{\left( {\mu \left( \lambda  \right)\dot x\left( \lambda  \right)} \right)} = eF\left( {x\left( \lambda  \right)} \right) \circ \dot x\left( \lambda  \right)\,,
\end{equation}
where
\begin{equation} \label{Faraday}
F_{ab}  \equiv \partial _a A^{\left(ext\right)}_b  - \partial _b A^{\left(ext\right)}_a {\text{,}}\quad \left( {F \circ \dot x} \right)_a  = F_{ab} \dot x^b\,.
\end{equation}
The mass can be inferred from the dynamics by forming the scalar product of Eq. (\ref{NewtonLorentz}) with ${\ddot x}$ and using the light speed constraint to give $\mu  = e{{\ddot x \circ \left( {F \circ \dot x} \right)} \mathord{\left/ {\vphantom {{\ddot x \circ \left( {F \circ \dot x} \right)} {\ddot x^2 }}} \right.
 \kern-\nulldelimiterspace} {\ddot x^2 }}$.

Eq. (\ref{NewtonLorentz}) is Lorentz covariant independent of the parameterization\footnote{It is sufficient to show that $\mu$ transforms like $d\lambda$  so that $d\lambda/\mu$ is a Lorentz scalar. From Eq. (\ref{NewtonLorentz}), and letting $\lambda  = f\left( \kappa  \right) \Rightarrow d\lambda  = \dot fd\kappa$, it follows in a few steps that $d\lambda/\mu$ is independent of $f$ and also therefore of however $\lambda$ may be affected by a transformation.} and invariant under simultaneous inversion of mass and charge. It describes only the passive response of the charge to an existing field however; the active - field-generation - behavior depends on the sign of the charge though not at all on the mass. A description of the 4-current and momentum that carries sufficient information for both active and passive roles must distinguish between the sign of the charge and the sign of the mass. Choosing again laboratory time $\lambda = x_0 \equiv t$ then $p  \equiv  {\left| \mu  \right|} \dot x$ defines a null 4-momentum with $p_0>0$ and with no loss of information in the spatial part ${\mathbf{p}} \in \mathbb{R}^3$ because $({\left| \mu  \right|,{\hat{\dot {\mathbf{x}}}}}) \in \mathbb{R}_ +   \times S^2  \approx \mathbb{R}^3$. A representation that preserves this definition and independently carries the information about the sign of the mass and having therefore the dimensionality of $( {\mu ,{\hat{\dot {\mathbf{x}}}}}) \in \R \times S^2$ - a double cover of $\mathbb{R}^3$ - can be given in terms of spinors.

%%%%%%%%%%%%%%%%%%%%%%%%%%%%%%%%%%%%%%%%%%%%%%%%%%%%%%%%%%%%%%%%%%%%%%%%%%%%%%%%%%%%%%%%%%%%%%%%%%%%%%%%%%%%%
\section{Remarks}
%%%%%%%%%%%%%%%%%%%%%%%%%%%%%%%%%%%%%%%%%%%%%%%%%%%%%%%%%%%%%%%%%%%%%%%%%%%%%%%%%%%%%%%%%%%%%%%%%%%%%%%%%%%%%
Direct action provides a useful framework in which to address the problem of infinite self-action of the classical charge. And, of relevance to that problem, it highlights ambiguities in, and naturally suggests alternatives to, the application of the Maxwell EM theory to charges at or near light speed. Two alternatives discussed here suggest a solution to the problem of infinite self-energy whilst reproducing the behavior of a classical charge in the sub-subluminal domain, i.e. as predicted by direct-action with a component of mechanical mass action. In the latter case the Maxwell equations remain valid and mass renormalization is required as usual. The modified actions differ in their predictions when the mechanical mass action is omitted. In that case, unlike in the traditional classical EM theory, the system predicts light-speed motion for a classical point charge with zero total self-energy in the absence of external fields.\footnote{Since EM has no intrinsic length scale the mass-scale must come from external interactions when intrinsic mechanical mass is excluded from the theory.} With these changes however the correspondence with the Maxwell field theory is destroyed.

So far the modified direct-action theories above have not been projected back into an equivalent field theory, though there are features suggestive of a connection with (first quantized) Dirac theory. Probably the best hope for this kind of approach is convergence with a de-Broglie-Bohm style formulation of the single-particle Dirac equation - as promoted for instance by \cite{Holland18} and \cite{Hestenes19} - wherein the null particle trajectories are the Bohmian flow lines.

%%%%%%%%%%%%%%%%%%%%%%%%%%%%%%%%%%%%%%%%%%%%%%%%%%%%%%%%%%%%%%%%%%%%%%%%%%%%%%%%%%%%%%%%%%%%%%%%%%%%%%%%%%%%%
\section*{Acknowledgements}
%%%%%%%%%%%%%%%%%%%%%%%%%%%%%%%%%%%%%%%%%%%%%%%%%%%%%%%%%%%%%%%%%%%%%%%%%%%%%%%%%%%%%%%%%%%%%%%%%%%%%%%%%%%%%
My thanks go to the founder of the PIRT series of conferences Michael Duffy, and to Peter Rowlands for his work to ensure a very enjoyable 2008 meeting at Imperial College. I am grateful to Eric Katerman for helpful discussions on the symmetry group of the action.

%%%%%%%%%%%%%%%%%%%%%%%%%%%%%%%%%%%%%%%%%%%%%%%%%%%%%%%%%%%%%%%%%%%%%%%%%%%%%%%%%%%%%%%%%%%%%%%%%%%%%%%%%%%%%

\end{document}